%% file: main.tex
\documentclass[alpha-refs]{wiley-article}

\usepackage{graphicx}
\usepackage[space]{grffile}
\usepackage{latexsym}
\usepackage{textcomp}
\usepackage{longtable}
\usepackage{tabulary}
\usepackage{booktabs,array,multirow}
\usepackage{amsfonts,amsmath,amssymb}
\usepackage{natbib}
\usepackage{url}
\usepackage{hyperref}
\hypersetup{colorlinks=false,pdfborder={0 0 0}}
\usepackage{etoolbox}
\newif\iflatexml\latexmlfalse

\AtBeginDocument{\DeclareGraphicsExtensions{.pdf,.PDF,.eps,.EPS,.png,.PNG,.tif,.TIF,.jpg,.JPG,.jpeg,.JPEG}}

\usepackage[utf8]{inputenc}
\usepackage[english]{babel}

\usepackage{siunitx}
\usepackage{overpic}
\usepackage{subfig}
\usepackage{caption}
\usepackage{amsfonts}
\usepackage{authblk}
\usepackage{orcidlink}

\iflatexml

\else
\fi

\usepackage{lineno}

\corraddress{Fabian Mockert, Institute of Meteorology and Climate Research Troposphere Research (IMKTRO), Karlsruhe Institute of Technology (KIT),PO 3640, 76021 Karlsruhe, Germany}
\corremail{fabian.mockert@kit.edu}
\presentadd{Institute of Meteorology and Climate Research Troposphere Research (IMKTRO), Karlsruhe Institute of Technology (KIT), PO 3640, 
76021 Karlsruhe, Germany}
\fundinginfo{European Union: ERC, ASPIRE, 101077260, Helmholtz Association: VH-NG-1243 Young Investigator Group 'SPREADOUT', KIT Centre MathSEE, Vector Stiftung: Young investigator Group 'Artificial Intelligence for Probabilistic Weather Forecasting', Axpo Solutions AG.}

\runningauthor{Mockert et al.}
\begin{document}

\include{0abstract}

\include{1introduction}

\include{2datamethods}

\include{3results}

\include{4discussionconclusion}

\section*{Conflict of interest}
All authors declare that they have no conflicts of interest.

\section*{Code and Data Availability}
Code is available on the GitHub repository \url{https://github.com/fmockert/postprocessingWR}. The ERA5 data can be obtained from the Climate Data Store \url{https://cds.climate.copernicus.eu/\#!/home}. Weather regime data are available from CMG upon request. The original S2S database is hosted at ECMWF as an extension of the TIGGE database.

\section*{Acknowledgements}\label{acknowledgements}
FM has received funding from the KIT Center for Mathematics in Sciences, Engineering and Economics under the seed funding programme. The contribution of JQ was partly funded by the European Union (ERC, ASPIRE, 101077260). The work of JQ and CMG was funded by the Helmholtz Association as part of the Young Investigator Group, “Sub-seasonal Predictability: Understanding the Role of Diabatic Outflow” (SPREADOUT, grant VH-NG-1243). SL gratefully acknowledges support by the Vector Stiftung through the Young Investigator Group “Artificial Intelligence for Probabilistic Weather Forecasting.” The contribution of MO was supported by Axpo Solutions AG

\bibliography{bibliography/library_PP.bib%
}
\newpage

\include{supplements}

\end{document}

%% file: 0abstract.tex
\title{\vspace{-1.1cm}Multivariate post-processing of probabilistic sub-seasonal weather regime forecasts\vspace{-0.8cm}}
\author[1\authfn{1}]{Fabian Mockert}
\author[1\authfn{2},2]{Christian M. Grams}
\author[3,4]{Sebastian Lerch}
\author[1\authfn{2},5,6,7]{Marisol Osman}
\author[1]{Julian Quinting}
\affil[1]{Institute of Meteorology and Climate Research Troposphere Research (IMKTRO), Karlsruhe Institute of Technology (KIT), Karlsruhe, Germany}
\affil[2]{now at: Federal Office of Meteorology and Climatology, MeteoSwiss, Zurich-Airport, Switzerland}
\affil[3]{Institute of Statistics, Karlsruhe Institute of Technology (KIT), Karlsruhe, Germany}
\affil[4]{Heidelberg Institute for Theoretical Studies, Heidelberg, Germany}
\affil[5]{now at: Universidad de Buenos Aires, Facultad de Ciencias Exactas y Naturales, Departamento de Ciencias de la Atmósfera y los Océanos, Buenos Aires, Argentina}
\affil[6]{now at: CONICET – Universidad de Buenos Aires. Centro de Investigaciones del Mar y la Atmósfera (CIMA), Buenos Aires, Argentina}
\affil[7]{now at: CNRS – IRD – CONICET – UBA. Instituto Franco-Argentino para el Estudio del Clima y sus Impactos (IRL 3351 IFAECI), Buenos Aires, Argentina}
\maketitle
\vspace{-0.4cm}
\begin{abstract}
\small
Reliable forecasts of quasi-stationary, recurrent, and persistent large-scale atmospheric circulation patterns -- so-called weather regimes -- are crucial for various socio-economic sectors, including energy, health, and agriculture. Despite steady progress, probabilistic weather regime predictions still exhibit biases in the exact timing and amplitude of weather regimes. This study thus aims at advancing probabilistic weather regime predictions in the North Atlantic-European region through ensemble post-processing. Here, we focus on the representation of seven year-round weather regimes in the sub-seasonal to seasonal reforecasts of the European Centre for Medium-Range Weather Forecasts (ECMWF). The manifestation of each of the seven regimes can be expressed by a continuous weather regime index, representing the projection of the instantaneous 500-hPa geopotential height anomalies (Z500A) onto the respective mean regime pattern. We apply a two-step ensemble post-processing involving first univariate ensemble model output statistics and second ensemble copula coupling, which restores the multivariate dependency structure. Compared to current forecast calibration practices, which rely on correcting the Z500 field by the lead time dependent mean bias, our approach extends the forecast skill horizon for daily/instantaneous regime forecasts moderately by 1.2 days to 14.5 days. Additionally, to our knowledge our study is the first to systematically evaluate the multivariate aspects of forecast quality for weather regime forecasts. Our method outperforms current practices in the multivariate aspect, as measured by the energy and variogram score. Still our study shows, that even with advanced post-processing weather regime prediction becomes difficult beyond 14 days, which likely points towards intrinsic limits of predictability for daily/instantaneous regime forecasts. The proposed method can easily be applied to operational weather regime forecasts, offering a neat alternative for cost- and time-efficient post-processing of real time weather regime forecasts.
\keywords{weather regimes, post-processing, ensemble model output statistics, ensemble copula coupling, forecasting}
\end{abstract}


%% file: 1introduction.tex
\section{Introduction}\label{Introduction}
Weather regimes, defined as quasi-stationary, recurrent, and persistent large-scale circulation patterns \citep{Michelangeli1995, Vautard1990}, provide valuable information for decision-making in the energy \citep{Bloomfield2021b, Mockert2023}, health \citep{Charlton-Perez2019}, and agricultural \citep{Lavaysse2018} sectors.
These weather regimes are associated with distinct conditions on surface variables, including 2-m temperature, 10-m wind, and radiation and thus prove to be beneficial for extended-range prediction. The representation of complex large-scale circulations through a finite set of states (e.g. weather regimes), facilitates the interpretation and categorisation of the prevailing large-scale circulation and its impact on surface weather. In the context of renewable energy forecasts for the European region, \citet{Bloomfield2021b} conducted a comprehensive study comparing grid-point-based forecasting methods with pattern-based methods (including weather regimes). While grid-point forecasts exhibit superior skill up to 10 days lead time, pattern-based methods demonstrate better performance at extended-range lead times (12+ days).

In this article, we adopt the year-round definition of seven North Atlantic-European weather regimes proposed by \citet{Grams2017}. These weather regimes, illustrated in the Supplement\,\ref{fig:WRpatterns}, represent large-scale circulation patterns within the 500 hPa geopotential height field (Z500). The regime definition roots in continuous information about the amplitude of the seven weather regimes via a normalised weather regime index (IWR). The seven dimensional IWR vector thus provides additional information about the current regime characteristics beyond a mere categorisation which proved to be useful in sub-seasonal prediction of weather regimes \citep[cf.\ discussion in][]{Grams2020}.

However, extended-range forecasts of Z500 have substantial biases \citep[e.g.][]{Ferranti2018, Bueler2021}. The current approach to address these mean biases involves calibrating the Z500 field before computing weather regimes \citep[cf.][]{Bueler2021}: Instead of computing the Z500 anomalies relative to the reanalysis climatology, this method computes anomalies relative to the 90-day running mean reforecast climatology at the respective lead time. While this correction mitigates the Z500 bias in the forecast field, it does not address all forecast errors, e.g. the flow dependence of forecasting errors. Additionally, this specific approach is impractical for operational on-the-fly reforecasts like those produced by the European Centre for Medium-Range Weather Forecasts (ECMWF) due to the need for compromises regarding averaging windows when computing running mean climatologies.

The presence of systematic errors and biases is a common challenge in ensemble weather forecasting \citep{Lerch2020, VannitsemEtAl2021}. Addressing this challenge can involve statistical post-processing methods, where systematic forecasting errors are corrected by analysing the statistical error distribution of past forecasts.
Most research efforts have been focused on univariate post-processing methods where different weather variables, locations or lead times are treated separately.
However, many practical applications require accurate representations of temporal, spatial and inter-variable dependencies. Preserving these multivariate dependency structures is crucial. In low-dimensional scenarios, this could be achieved by fitting a specific multivariate probability distribution \citep{Schefzik2018}. A more broadly applicable strategy involves a two-step process. In the first step, forecasting variables are individually post-processed in all dimensions. Then, in the second step, the multivariate dependency structure is restored by re-arranging univariate sample values based on the rank order structure of a specific multivariate dependence template. From a mathematical perspective, this second step corresponds to applying a parametric or non-parametric copula. Commonly used multivariate approaches include ensemble copula coupling \citep{Schefzik2013}, the Schaake-Shuffle \citep{Clark2004} or a Gaussian copula approach \citep{Moeller2013}.

The combination of ensemble model output statistics \citep[EMOS,][]{Gneiting2007a} and ensemble copula coupling \citep[ECC,][]{Schefzik2013} grounded in Sklar's theorem from multivariate statistics \citep{Schefzik2013}, EMOS-ECC, has proven to be effective across various meteorological forecast variables. \citet{Schefzik2013} explored different EMOS-ECC configurations for sea level pressure forecasts and consistently found its superior performance compared to alternative calibration techniques. Additionally, the configuration where equidistant quantiles are drawn from the forecast distribution (EMOS-ECC-Q) outperformed other EMOS-ECC configurations in the context of pressure forecasts. In a study by \citet{Scheuerer2015} on wind speed forecasts, EMOS-ECC-Q consistently outperformed non-calibrated ensemble forecasts, EMOS-Q (without restoring multivariate dependencies), and another EMOS-ECC configuration. Applying the EMOS-ECC approach from \citet{Schefzik2013} to temperature forecasts, \citet{Schefzik2017} compared it with a member-by-member post-processing (MBMP) method. Their findings indicate that both methods exhibit good performance, with EMOS-ECC consistently outperforming MBMP in predictive skill. 
Further, various comparative studies of multivariate post-processing methods have found that the differences between different variants of ECC or observation-based approaches such as the Schaake shuffle, tend to be small and that EMOS-ECC-Q usually constitutes a competitive benchmark \citep[e.g.,][]{Wilks2015,Lerch2020,Perrone2020,Lakatos2023}.
Commonly, as also in the studies mentioned above, post-processing is applied directly to meteorological forecast variables, e.g.\ temperature, precipitation or wind speed forecasts. The promising results from these recent studies motivate us to apply the EMOS-ECC-Q approach to our problem at hand, enabling comprehensive multivariate post-processing. Instead of directly post-processing a forecast variable (e.g., Z500) generated by the forecasting model, we post-process the weather regime index forecast, which is derived from the Z500 forecast.

In our study, we develop a two-step post-processing method that adapts EMOS-ECC to forecasts of the continuous weather regime index IWR. 
In a first step, we address the univariate marginal distributions of each weather regime index independently. Subsequently, in the second step, we restore the multivariate dependency structure among weather regimes by applying a copula function. This function learns the dependence structure from the non-calibrated weather regime index forecasts. To our knowledge, the weather regime index has so far not been calibrated on a multivariate level. Our study marks the first comprehensive evaluation of the probabilistic forecast skill for weather regime forecasts at a multivariate level.

The structure of the article is outlined as follows: In Section\,\ref{sec:DataMethod}, we introduce the reforecasts, the weather regime index and the statistical post-processing methods. Additionally, we discuss the scoring rules used to evaluate the forecasts. Section\,\ref{sec:Results} starts with a brief discussion of the mean biases in ECMWF Integrated forecasting system (IFS) reforecasts and is then divided into 3 parts. In Section\,\ref{sec:univariate}, we analyse the univariate skill. Then we shift our focus to assess multivariate skill (Section\,\ref{sec:multivariate}) and lastly, we test the sensitivity of the EMOS-ECC approach on the frequency of reforecast initialisations and historical period covered by the reforecasts (Section\,\ref{sec:operational}). In Section\,\ref{sec:Discussion}, we conclude and discuss our findings, and give an outlook on further research avenues.

%% file: 2datamethods.tex
\section{Data and methods}\label{sec:DataMethod}
\subsection{ECMWF reforecast and reanalysis}
For our study, we utilise sub-seasonal to seasonal reforecast data by the European Centre for Medium-Range Weather Forecasts (ECMWF), provided through the Subseasonal-to-Seasonal (S2S) Prediction Project Database \citep{Vitart2017}. To increase the number of forecast initial dates available for our analysis, we merge forecasts from two consecutive model cycles, Cy46R1 and Cy47R1 \citep{ECMWF2018}. These reforecasts are computed twice a week (Mondays and Thursdays) and consist of 11 ensemble members, covering a forecast lead time of 0--46 days with 91 vertical levels and a native horizontal grid spacing of 16\,km up to day 15 and 32\,km from day 15 onwards. Forecast data were remapped from their native resolution to a regular latitude-longitude grid with 1\textdegree grid spacing.
The two model cycles were operational from June 11, 2019, to May 11, 2021, with a cycle change on June 30, 2020. As a result, for the period between May 11 and June 11 reforecasts are only available from Cy46R1, which may impact the training and evaluation of our methods within that specific time period. It is likely that in this period, the post-processing performs worse due to less training dates. Nonetheless, with the combination of Cy46R1 and Cy47R1, our dataset spans 21 years of reforecasts, from June 11, 1999 to May 11, 2020 comprising a total of 4000 initial dates.

The reforecast are initialised using the ERA5 reanalysis data. Therefore, for verification purposes, we employ the ERA5 dataset \citep{Hersbach2020} remapped to the same grid spacing as our ground truth.

\subsection{Weather regimes}
In this study, we use the seven year-round North Atlantic-European weather regimes introduced by \citet{Grams2017} based on ERA-Interim reanalysis but adapted here for the newer ERA5 reanalysis as described in \citet{Hauser2023, Hauser2023pre} and applied to IFS reforecasts following the approach of \citet{Bueler2021} and \citet{Osman2023}. These weather regimes represent the most common large-scale circulation patterns in the North Atlantic-European region (30--90\textdegree N, 80\textdegree W--40\textdegree E). In brief, we conduct an empirical orthogonal function (EOF) analysis of 6\,hourly (1979-2019), 10-day low pass filtered (filter-width of 20\,days, hence $\pm10$\,days), seasonally normalised geopotential height anomalies (Z500A, relative to 91-day running mean climatology) within the domain of the weather regimes. We then apply a k-means clustering algorithm on the first seven EOFs, which yields an optimal number of seven cluster means representing the seven distinct weather regimes (Figure\,\ref{fig:WRpatterns}), with three cyclonic (Atlantic Trough (AT), Zonal Regime (ZO), and Scandinavian Trough (ScTr)) and four anticyclonic regime types (Atlantic Ridge (AR), European Blocking (EuBL), Scandinavian Blocking (ScBL), and Greenland Blocking (GL)).

To describe the projection of instantaneous anomalies on the mean regime patterns, whether in reanalysis or (re-)forecast, we introduce a seven-dimensional weather regime index (IWR). We briefly outline the steps to compute the weather regime index (see also Figure\,\ref{fig:arrowdiagram}) for the non-calibrated forecasts and refer to \citet{Bueler2021} and \citet{Osman2023} for a more detailed description on the computation of the IWR.
We compute the Z500A relative to the 91-day running mean climatology. Similar to the definition of the weather regime patterns, we then apply a low-pass filter and normalisation to the Z500A. These standardised and filtered Z500A are projected onto the seven cluster mean Z500A following the method of \citet{Michel2011} via
\begin{equation}
P_{(wr)}(t,\tau) = \frac{1}{\sum_{\lambda,\varphi\in(region)}\cos(\varphi)} \sum_{\lambda,\varphi\in(region)}\left(\Phi(\lambda,\varphi,t,\tau) \cdot \Phi_{(wr)}(\lambda,\varphi)\cdot\cos(\varphi)\right).
\end{equation}
Here, $P_{(wr)}(t,\tau)$ represents a scalar measure for the spatial correlation of the instantaneous anomaly field $\Phi(\lambda,\varphi,t,\tau)$ at lead time day $\tau$ for a specific initialisation date $t$ and ensemble member to the cluster mean anomaly field $\Phi_{(wr)}(\lambda,\varphi)$ for the weather regimes $wr\in\{AT, ZO, ScTr, AR, EuBL, ScBL, GL\}$. $\lambda$ and $\varphi$ denote the longitudinal and latitudinal degrees, respectively.

The weather regime index $I_{(wr)}(t,\tau)$ for each weather regime, lead time, and ensemble member is then computed based on the anomalies of the projections $P_{(wr)}(t,\tau)$ with respect to the climatological mean projection $\overline{P_{(wr)}} = \frac{1}{N} \sum_{t\in IC_{train}} P_{(wr)}(t,\tau)$, with $IC_{train}$ the initialisation dates in the training period from 1999--2015, and N the number of initialisation dates in $IC_{train}$,
\begin{equation}
I_{(wr)}(t,\tau) = \frac{P_{(wr)}(t,\tau)-\overline{P_{(wr)}}}{\sqrt{\frac{1}{N-1}\sum_{t\in IC_{train}}\left( P_{(wr)}(t,\tau)-\overline{P_{(wr)}}\right)^2}},
\end{equation} 
where the denominator represents the climatological standard deviation of the projection.
Combining the weather regime index of each weather regime, we obtain a 7-dimensional weather regime index vector $IWR(t,\tau)$ for each initialisation date plus lead time, and ensemble member.
Computing the Z500A by simply taking the difference between the model (re)forecast and ERA5 climatology (June 1999-- May2015, excluding June 2015--May 2020 to reserve it for the use as a testing set later on) does not account for systematic model biases, making it non-calibrated. Consequently, we refer to these weather regime indices as raw or non-calibrated (abbreviated as non cal.) forecasts. To address this issue and eliminate systematic Z500 forecast biases from the anomalies, we replace the underlying Z500 calendar day climatology derived from the ERA5 reanalysis with a model climatology \citep[similar to][]{Bueler2021}. We refer to this adapted method as Z500 calibrated forecast (abbreviated as Z500 cal.), which serves, alongside the non-calibrated forecasts, as our baseline against which we compare our statistical post-processing approach. To ensure a fair comparison later on, we again divide our 21 years of forecasts into a training set spanning from June 1999 to May 2015 and a testing set covering June 2015 to May 2020. The Z500 model climatology is computed based on the training dataset.

When comparing the forecasts of the weather regime index to a climatological reference forecast, we always refer to the climatology from the ERA5 perfect member obtained from the raw, non-calibrated forecast. 

\subsection{Statistical post-processing}
We employ a two-step statistical post-processing method based on Sklar's theorem \citep{Sklar1959, Schefzik2013, Lerch2020} for the seven dimensional weather regime index forecast. First, univariate processing using ensemble model output statistics (EMOS \citep{Gneiting2007a}), and second, multivariate processing through ensemble copula coupling (ECC \citep{Schefzik2013}). We adapt the notation of \citet{Lerch2020} and \citet{Chen2022} with slight modifications to our specific setup.

According to Sklar's theorem, a multivariate cumulative distribution function (CDF) $H$ can be decomposed into a copula function $C$ representing the dependence structures and its marginal univariate CDFs $F_{AT}, ..., F_{GL}$ obtained through univariate post-processing. Specifically, for $x_{AT}, ..., x_{GL} \in \mathbb{R}$, we have $H(x_{AT}, ..., x_{GL}) = C(F_{AT}(x_{AT}), ..., F_{GL}(x_{GL}))$ \citep{Lerch2020}, where the subscript iterates through the 7 weather regimes $wr \in \{AT, ZO, ScTr, AR, EuBL, ScBL, GL\}$, and $x$ represents the weather regime index.

The unprocessed 7D ensemble forecast of the weather regime index with $M=11$ ensemble members is denoted as $\textbf{X}_1, ..., \textbf{X}_M \in \mathbb{R}^7$, where $\textbf{X}_m = \left( X_m^{(AT)}, ..., X_m^{(GL)} \right)$. Similarly, the observations of the weather regime index are denoted as $\textbf{y} = \left( y^{(AT)}, ..., y^{(GL)} \right) \in \mathbb{R}^7$.
\subsubsection{Ensemble model output statistics}
The first step of the two-step post-processing method is to univariately apply EMOS to fit a Gaussian predictive distribution $\mathcal{N}^{(wr)}$ with mean $\mu$ and variance $\sigma^2$,
\begin{equation}
    y^{(wr)} | X_1^{(wr)}, ..., X_M^{(wr)} \sim \mathcal{N}(\mu, \sigma^2) = F_{\theta}^{(wr)},
\end{equation} 
where the distribution parameters $\theta = (\mu,\sigma)$ are linked to the ensemble forecasts via $\textbf{$\theta$} = g(X_1, ..., X_M)$. 
\begin{figure}[!h] 
    \centering
    \includegraphics[width=1\linewidth]{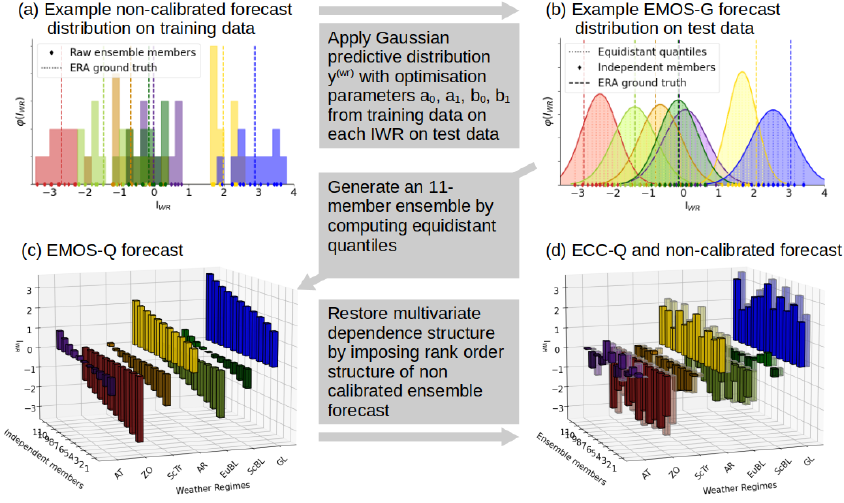}
    \caption{Post-processing workflow for the seven dimensional weather regime index using EMOS-ECC. (a) Initially, the optimisation paramaters are fitted by minimising the continuous ranked probability score on the non-calibrated ensemble forecasts in the training data. The obtained optimising parameters from the training data are then applied on ensemble forecasts in the test data to transform them into Gaussian predictive distributions (a to b). (b) The resulting EMOS-G forecast is used to sample equidistant quantiles, generating (c) the post-processed ensemble forecast EMOS-Q, which is organised by the rank. Finally, (d) ECC-Q is employed to restore the multivariate dependency structure of the non-calibrated forecast (transparent bars in (d)).}
    \label{fig:ppworkflow}
\end{figure}
The parameters $(\mu, \sigma^2) = (a_0 + a_1\overline{X}, b_0 + b_1 S^2) = g\left(X_1^{(wr)}, ...,X_M^{(wr)}\right)$ are determined by minimising the continuous ranked probability score (CRPS) via optimisation of $a_0, a_1, b_0, b_1$ on a training period from June 1999 to May 2015 (Figure\,\ref{fig:ppworkflow}a, b). The resulting model for each weather regime index is referred to as EMOS-G. Initially, we conducted the above mentioned univariate post-processing step using different setups for the training period to discern variations in performance. These setups involved training one EMOS method on the full dataset, splitting it into two seasons - winter half-year (October to March) and summer half-year (April to September), four seasons - winter (December to February), spring (March to May), summer (June to August), and autumn (September to November), or training EMOS for each calendar day using running windows of 9, 31, and 91 days. The performance of all methods was comparable, with a slight tendency towards better performance for the four-season approach and a running window of 31 days. We, therefore, focus our analysis on the 31-day running window setup, as it is amongst the best performing setup and compatible with on-the-fly generated reforecast data.

For each weather regime index, we reduce the continuous Gaussian forecast distributions to an ensemble with the same number of members ($M=11$) as the unprocessed forecast $\textbf{X}_1, ..., \textbf{X}_M$ (Figure\,\ref{fig:ppworkflow}b). This process involves drawing equidistant quantiles at levels $\frac{1}{M+1}, ..., \frac{M}{M+1}$ from the forecast distributions $F_\theta^{(wr)}$ and the resulting forecasts are referred to as EMOS-Q (Figure\,\ref{fig:ppworkflow}c).

\subsubsection{Ensemble copula coupling}
In the second step, we utilise ECC to restore the multivariate dependency structure of the weather regime index. We achieve this by retrieving the rank order structure of the unprocessed ensemble member forecasts  (Figure\,\ref{fig:ppworkflow}d, pale bars) and sorting the post-processed ensemble members accordingly (Figure\,\ref{fig:ppworkflow}c to\,\ref{fig:ppworkflow}d). To formalise this process, for each weather regime index, we define $\sigma_{(wr)}(m) = \textrm{rank}\left( X_m^{(wr)} \right)$ as a permutation, and aim to find $\Tilde{X}_m^{(wr)} = \hat{x}_{\sigma_{(wr)}(m)}^{(wr)}$, where $\hat{x}_1^{(wr)}, ..., \hat{x}_M^{(wr)}$ form a sample under the assumption that $\hat{x}_1^{(wr)} \leq ... \leq \hat{x}_M^{(wr)}$ to simplify the notation. Here, $\hat{x}_1^{(wr)} := \left( F_\theta^{(wr)} \right)^{-1}\left(\frac{1}{M+1}\right), ..., \hat{x}_M^{(wr)} := \left( F_\theta^{(wr)} \right)^{-1}\left(\frac{M}{M+1}\right)$ represent the quantiled EMOS-Q weather regime indices. This non-parametric, empirical copula approach is referred to as ECC-Q.
Unlike in the introduction, for the sake of simplicity, we will refer to the execution of the two-step process as ECC and not EMOS-ECC since we did not test any other univariate methods. Next to ECC, we also tested the Schaake Shuffle approach, which ranks ensemble members based on past observations rather than the actual forecast.

\subsection{Skill metrics and their skill scores}
In this study, we aim to compare the univariate and multivariate skill of the post-processed ensemble forecasts using EMOS-G and ECC-Q with the skill of the current practice of processing the Z500 field prior to computing the weather regime index. To conduct this comparison, we introduce the continuous ranked probability skill score (CRPSS) for univariate evaluation, and the energy skill score (ESS) and variogram skill score (VSS) for assessing multivariate aspects of forecast quality. While these scores are extensively discussed in \citet{Gneiting2007a} and \citet{Scheuerer2015}, we will provide a concise summary of the underlying metrics and skill scores in this paper and direct interested readers to the mentioned literature for more detailed information.

The continuous ranked probability score (CRPS) is a metric used for evaluating univariate probabilistic forecasts and generalises the absolute error to which it reduces when the forecast is deterministic. It is defined as:
\begin{equation}
CRPS(F,y) = \int_{-\infty}^{\infty} \left( F(x) - \mathbb{1}\{ x \geq y \} \right)^2\textrm{d}x,
\end{equation}
where $\mathbb{1}$ represents the indicator function, $F$ is a predictive distribution, and $y$ is the observation. Lower values represent higher skill where a CRPS value of 0 indicates a perfect forecast.

To assess the skill of multivariate probabilistic forecasts, we employ the Energy Score (ES) and the Variogram Score of order $p$ (VS$^p$). The ES is a generalisation of the CRPS and the variogram score origins in the concept of variograms (also referred as structure functions) from geostatistics.

The ES is calculated as:
\begin{equation}
ES(F,\textbf{y}) = \frac{1}{M}\sum_{i=1}^M \left\| \textbf{X}_i - \textbf{y} \right\| - \frac{1}{2M^2}\sum_{i=1}^M\sum_{j=1}^M \left\|\textbf{X}_i - \textbf{X}_j\right\|,
\end{equation}
where $\|\cdot\|$ denotes the Euclidean norm on $\mathbb{R}^D$, $\textbf{X}_i,\textbf{X}_j$ are samples from the multivariate forecast distribution, and $\textbf{y}$ represents the multivariate observation, respectively.

The VS$^p$ is given by:
\begin{equation}
VS^p(F,\textbf{y}) = \sum_{i=1}^D\sum_{j=1}^D w_{i,j}\left( \left| y^{(i)} - y^{(j)} \right|^p -\frac{1}{M}\sum_{k=1}^M \left| X_k^{(i)} - X_k^{(j)} \right|^p \right)^2,
\end{equation}
with $w_{i,j}$ being a non-negative weight for pairs of component combinations, and $p$ representing the order of the VS. In accordance with \cite{Scheuerer2015}, we use an unweighted version of the VS with $w_{i,j}=1$ and set $p=0.5$. Both, the values of the energy and variogram score can be interpreted similar to the CRPS, where lower values represent better forecast skill.

The skill scores ($SS_f$) for the mentioned skill metrics ($S\in\{CRPS, ES, VS^p\}$) are calculated as 
\begin{equation}
SS_f = \frac{\overline{S_{ref}}-\overline{S_{f}}}{\overline{S_{ref}}-\overline{S_{opt}}} = 1-\frac{\overline{S_{f}}}{\overline{S_{ref}}},
\end{equation}
where $\overline{S_{f}}$ is the mean score of a forecasting method $f$. For the skill metrics considered here, perfect forecasts receive a score value of 0, i.e., $\overline{S_{opt}}=0$. As a reference forecast ($\overline{S_{ref}}$), we generally use a 31-day ensemble climatology, where the ensemble members are represented by reanalysis data for the S2S reforecast dates throughout the training period of June 1999 to May 2015 of the non-calibrated weather regime index. A value of 1 indicates perfect skill, a value of 0 equal skill and negative values worse skill than the reference forecast. 

\subsection{Diebold-Mariano test of equal performance}
To assess the statistical significance of the differences in predictive performance between the post-processed forecast, non-calibrated forecast, Z500 calibrated forecasts, and climatological reference forecasts, we employ the Diebold-Mariano test of equal performance \citep{Diebold1995}, the test statistic $t_n$ of which is given by 
\begin{equation}
t_n = \sqrt{n} \frac{\overline{S_n^F}-\overline{S_n^G}}{\hat{\sigma}_n},
\end{equation}
where $\overline{S_n^F} = \frac{1}{n}\sum_{i=1}^n S(F_i, \textbf{y}_i)$ and $\overline{S_n^G} = \frac{1}{n}\sum_{i=1}^n S(G_i, \textbf{y}_i)$ represent the mean scores of forecasts $F$ and $G$ over $n$ samples, respectively, and $\hat{\sigma}_n = \sqrt{\frac{1}{n}\sum_{i=1}^n\left( S(F_i, \textbf{y}_i) - S(G_i, \textbf{y}_i) \right)^2}$ is the sample standard deviation. 
Under standard regularity assumptions, $t_n$ asymptotically follows a standard Gaussian distribution. 
Negative values of $t_n$ indicate that $F$ outperforms $G$ with respect to the considered score $S$. We use a level of $\alpha=0.05$ to assess the significance of the performance. Values falling outside this level (indicated with grey shading in Figure\,\ref{fig:Diebold_univariate} and Figure\,\ref{fig:Diebold_multivariate}) are considered statistically significant.

\subsection{Verification rank histograms}
Verification rank histograms are essential tools for assessing the calibration of a collection of ensemble forecasts for a scalar predictand which is here the IWR \citep{Wilks82011}. To construct a rank histogram, we analyse $n$ ensemble forecasts, each with $n_{ens}=11$ ensemble members. For every ensemble forecast, we determine the rank of the observation within the $n_{ens}+1$ values, hence we sort the IWR in ensemble forecasts and observation (reanalysis) in ascending order and determine the rank of the observation value. These ranks are then tabulated, resulting in a histogram that represents the distribution of observation ranks across all ensemble forecasts. A calibrated and reliable forecast would be represented by a uniform distribution of ranks, while deviations may indicate biases or under-/overdispersion in the ensemble forecasts. To provide a comprehensive view, we visualise the verification rank histograms of each lead time in one joint 2-d histogram. The rank is on the x-axis, the lead time on the y-axis and the frequency distribution is indicated by coloured boxes and numbers inside the boxes, indicating the deviations from a perfect frequency distribution of 1/12 for each rank due to 11+1 members.

%% file: 3results.tex
\section{Results}\label{sec:Results}
In this section, we first analyse Z500 forecast biases and how they are connected to biases in the weather regime index forecasts. Then, we present a comprehensive evaluation of the post-processed forecasts in comparison to both the non-calibrated and Z500 calibrated forecasts of the weather regime index (IWR). Our analysis is divided into two main aspects: the assessment of univariate skill, focusing on EMOS-G (Section\,\ref{sec:univariate}), and the evaluation of multivariate skill, centred around ECC-Q (Section\,\ref{sec:multivariate}). In addition, we investigate the method's sensitivity to variations in training data availability (Section\,\ref{sec:operational}). It is important to note that we are assessing the ensemble forecast's capability to predict IWR on given days, which is a challenging forecasting question, in particular at extended-range lead times.
\subsection{Z500 forecast biases}\label{sec:Z500biases}
\begin{figure}[!ht]   
\includegraphics[width=1\textwidth]{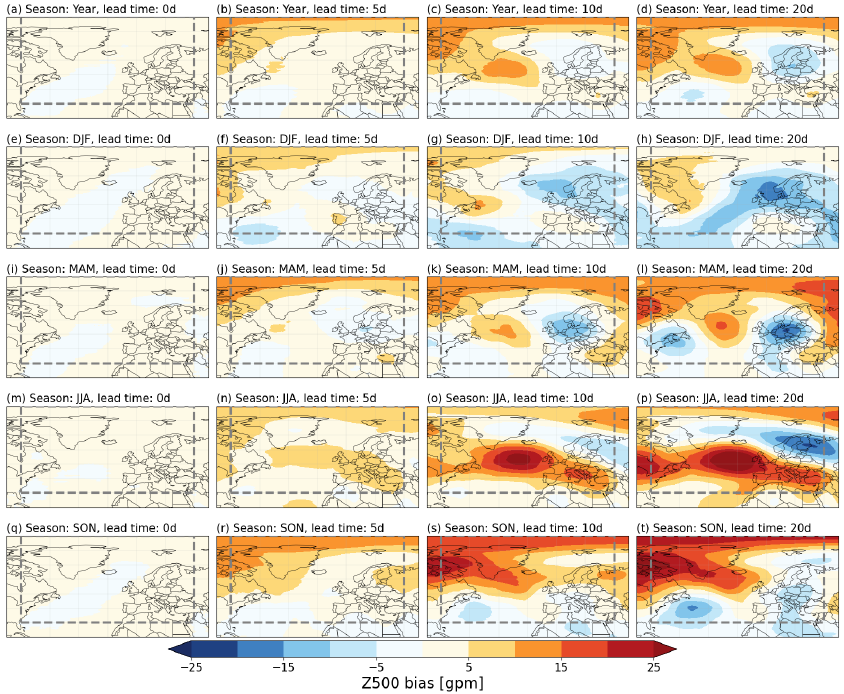}
\caption{Displayed are the mean 500 hPa geopotential height anomaly biases of the ensemble mean forecasts. The Z500A bias is computed by subtracting the ERA perfect member from the ensemble mean. The mean bias fields are presented for the entire year and four seasons (winter: DJF, spring: MAM, summer: JJA, autumn: SON) across rows, and at lead times of 0, 5, 10, and 20 days across columns. The region corresponding to North Atlantic-European weather regimes is indicated by the dashed grey box.}
\label{fig:Z500bias}
\end{figure}
Biases in the Z500 forecast have a direct impact on the seven dimensional weather regime index as these directly project into the IWR forecasts. The analysis of Z500 biases as a function of forecast lead time (Figure\,\ref{fig:Z500bias}) unveils the reasons for systematic biases in the IWR (Figure\,\ref{fig:IWRbias}). Forecast biases grow from values near 0\,gpm at 0 day lead time to around 40\,gpm at lead times beyond 20 days. Throughout the year, positive biases dominate in the northern part of the weather regime region (denoted by the grey dashed box in Figure\,\ref{fig:Z500bias}), and extend from Canada into the high-latitude North Atlantic. This positive bias anomaly most prominently projects into the Atlantic Ridge weather regime (yellow in Figure\,\ref{fig:IWRbias}).
The Z500 biases are seasonally dependent which also manifest in the IWR. In winter (second row in Figure\,\ref{fig:Z500bias}), Z500 biases exhibit a dipole structure projecting into the Atlantic Trough and Greenland Blocking (purple and blue in Figure\,\ref{fig:IWRbias}). In spring (third row in Figure\,\ref{fig:Z500bias}), the positive Z500A in the North Atlantic grows, projecting into the Atlantic Ridge and Greenland Blocking (yellow and blue in Figure\,\ref{fig:IWRbias}). During summer (fourth row in Figure\,\ref{fig:Z500bias}), the positive anomaly in the North Atlantic intensifies, accompanied by a negative anomaly in northern Europe, projecting into the Scandinavian Trough and Atlantic Ridge (orange and yellow in Figure\,\ref{fig:IWRbias}). In autumn (fifth row in Figure\,\ref{fig:Z500bias}), the positive anomaly resides mainly in northern America and the high-latitude North Atlantic, resulting in projections into the Greenland Blocking and Atlantic Ridge (blue and yellow in Figure\,\ref{fig:IWRbias}). This seasonally differentiated analysis of bias growth in the Z500 field and its projection into weather regimes emphasises the need for bias correction.

\subsection{Univariate post-processing}\label{sec:univariate} 
Here, we utilise EMOS for univariate post-processing. EMOS is trained on a reforecast dataset with 3101 forecasts spanning June 1999 to May 2015, and subsequently tested on 899 forecasts from the reforecast dataset covering June 2015 to May 2020.
As we intend to apply the post-processing to operational forecasts, we train the EMOS for each calendar day using a running window approach with 31 days and non-calibrated forecasts.

Upon comparison, including the full information provided by the continuous Gaussian distribution (EMOS-G) exhibits slightly higher skill than the post-processed ensembles generated by drawing equidistant quantiles (EMOS-Q), albeit the difference being marginal. Consequently, we adopt EMOS-G as the superior method for univariate post-processing and evaluate the CRPS of EMOS-G instead of EMOS-Q. However, for assessing the calibration and reliability of forecasts, we utilise verification rank histograms for all forecasting methods. Accordingly, we rather use the quantiled distributions by EMOS-Q forecasts than comparing Probability Integral Transform histograms (PIT histograms) for EMOS-G forecasts with verification rank histograms for the discrete non-calibrated and Z500 calibrated forecasts.

In the previous section, we identified biases in the Z500 field forecasts (Figure\,\ref{fig:Z500bias}). The (positive) bias is particularly prominent in the high latitudes of the North Atlantic, persisting across seasons and lead times, notably projecting into the Atlantic Ridge regime which is characterised by a positive Z500A in a similar region. Therefore, we focus on analysing the verification rank histograms for the Atlantic Ridge and its counterpart, the Atlantic Trough, aiming to assess forecast reliability and calibration. For a comprehensive overview of rank histograms for all weather regimes, please refer to Figures\,\ref{fig:2D_rankhistcyclonic} and\,\ref{fig:2D_rankhistblocked} in the Supplement.
\begin{figure}[!ht]   
\includegraphics[width=1\textwidth]{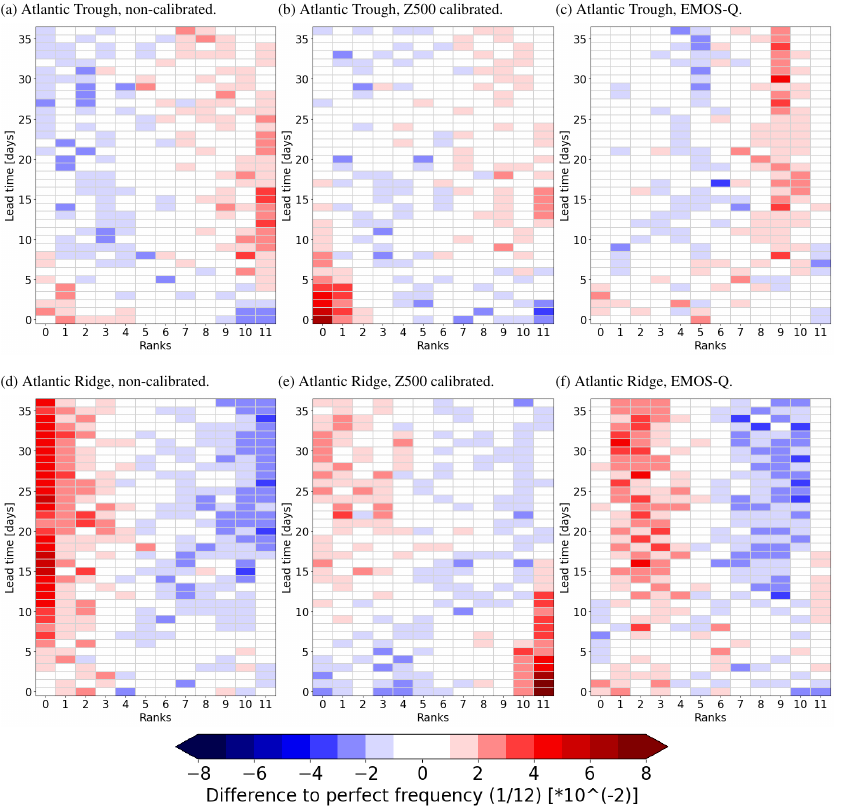}
\caption{The verification rank histograms for the Atlantic Trough (a-c) and the Atlantic Ridge (d-f) are visualised. Additional rank histograms for all cyclonic and anticyclonic weather regimes can be found in supplementary Figures\,\ref{fig:2D_rankhistcyclonic} and\,\ref{fig:2D_rankhistblocked}. The figures illustrate the rank histograms for the non-calibrated forecasts (a, d), Z500 calibrated forecasts (b, e), and EMOS-Q post-processed forecasts (c, f) for both weather regimes. The rank is illustrated on the x-axis, the lead time on the y-axis and the frequency of occurrence, specific for each lead time, is indicated by the values and colours of the boxes. To facilitate the readability, the frequency is shown as anomaly to a perfect distribution of 1/12.}
\label{fig:2D_rankhistexamples}
\end{figure}

Due to the observed bias in the Z500 forecast, the non-calibrated IWR forecast for Atlantic Trough tends to be underforecast for lead times exceeding 10 days (Figure\,\ref{fig:2D_rankhistexamples}a, noticeable in red colours at higher ranks), while overforecast on shorter lead times. Similarly, the Atlantic Ridge is consistently overforecasted across all lead times (Figure\,\ref{fig:2D_rankhistexamples}d, evident in red colour at lower ranks). Although the bias correction of the Z500 field improves the reliability for these two weather regimes on extended lead times (Figure\,\ref{fig:2D_rankhistexamples}b and Figure\,\ref{fig:2D_rankhistexamples}e), it deteriorates the reliability on shorter lead times. Forecasts up to 10 day lead time tend to overforecast the Atlantic Trough regime, while the Atlantic Ridge is strongly underforecasted on lead times up to 12 days after the Z500 bias correction. 
Verification rank histograms show that forecasts post-processed with EMOS exhibit improved reliability (Figure\,\ref{fig:2D_rankhistexamples}c and Figure\,\ref{fig:2D_rankhistexamples}f for the Atlantic Trough and Atlantic Ridge, respectively). While forecast biases persist to some extent, their magnitude decreases, and the largest deviations shift from the outer ranks towards central ranks. This general trend is also observed across the other weather regimes. EMOS consistently enhances the non-calibrated IWR forecast, particularly for the Zonal Regime and Scandinavian Trough (see Supplementary Figure\,\ref{fig:2D_rankhistcyclonic}). Forecasts are also better calibrated for Greenland Blocking. However, forecasts of European and Scandinavian Blocking exhibit similar or even larger miscalibration, with observation too frequently falling into the highest ranks (see Figure\,\ref{fig:2D_rankhistblocked}). 
In summary, the Z500 calibrated forecast, especially on shorter lead times, degrades the ensemble's calibration. The IWR is either underforecasted (Scandinavian Trough), overforecasted (Greenland Blocking), or overconfident (Zonal Regime, European and Scandinavian Blocking), as evident in the underdispersive distribution in the verification rank histograms. Overall, these findings suggests that EMOS post-processing generates more reliable and consistent forecasts for the weather regime index compared to Z500 calibrated forecasts.

\begin{figure}[!ht]   
\includegraphics[width=1\textwidth]{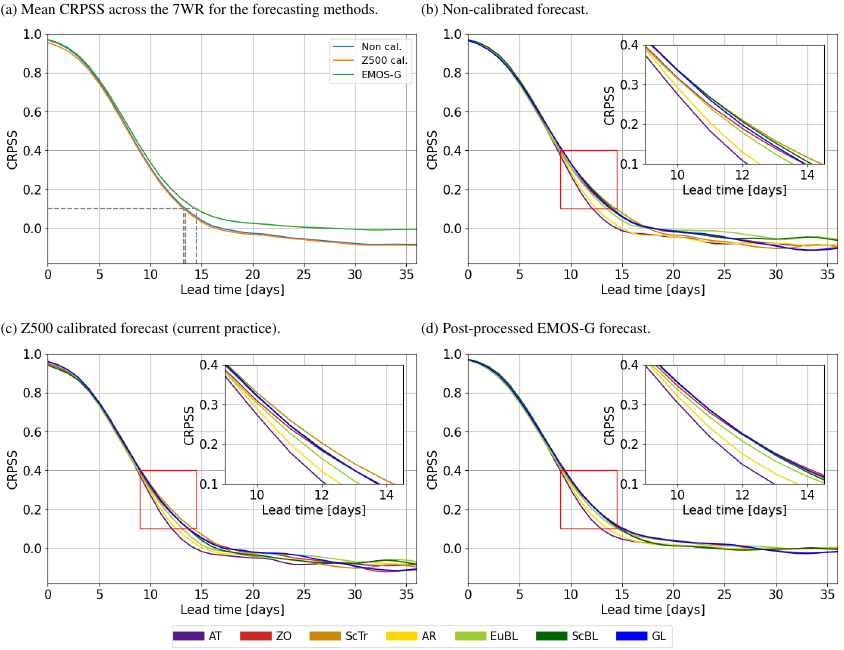}
\caption{CRPSS as a function of lead time. In (a), the aggregated mean CRPSS across the weather regimes is shown for the three forecasting methods, non-calibrated (blue), Z500 calibrated (orange) and EMOS-G (green). In (b)--(d) the CRPSS is shown separately for each weather regime, indicated by the regime colours, for (b) the non-calibrated forecasts, (c) the Z500 calibrated forecasts and (d) the EMOS-G forecasts. The red box indicates the area in which we zoom in to better visualise the differences between those regimes. The CRPSS is calculated against a 31-day rolling climatological forecast.}
\label{fig:CRPSS}
\end{figure}

We now dive deeper into the comparison of the univariate skill of the various forecasting methods (non-calibrated, Z500 calibrated, and EMOS-G). For this purpose, we assess the continuous ranked probability skill score (CRPSS) with climatology as the reference forecast. As an initial comparison among the forecasting methods, we examine the mean skill score over the 7 weather regimes (Figure\,\ref{fig:CRPSS}a). 

The mean skill scores of the three forecasting methods (Figure\,\ref{fig:CRPSS}a) mainly differ for extended-range lead times with the CRPSS for EMOS-G approaching 0 and only surpassing it minimally at lead times beyond 30 days. The CRPSS for the non-calibrated and Z500 calibrated forecasts is greater than 0 until day 17 and approaches a skill score of --0.09. For early lead times, all forecasting methods obtain high scores, with the non-calibrated and EMOS-G forecast slightly outperforming the Z500 calibrated forecasts. On the extended lead times, the Z500 calibrated forecasts exhibit slightly lower CRPSS values than the non-calibrated forecasts, though the difference is minimal.

In Figure\,\ref{fig:CRPSS}a, we observed that the mean skill scores of the non-calibrated and EMOS-G forecast remain similar up to day 10 and afterwards EMOS-G forecasts show noticeably higher skill than the non-calibrated and Z500 calibrated forecasts. We now analyse the skill scores across the weather regimes (Figures\,\ref{fig:CRPSS}b-\ref{fig:CRPSS}d) to reveal commonalities and differences between the individual forecasting methods.
All three forecasting methods have in common that the difference in the CRPSS of the individual weather regimes are close to indistinguishable up to a lead time of 7 days. A further commonality is the order of skill for the different weather regimes on lead times between 10 and 15 days (the lead time range until where all forecasting methods still exhibit CRPSS values larger than 0 for each weather regime). The lowest skill is observed for Atlantic Trough and Atlantic Ridge, which coincides with our choice in the analysis of verification rank histograms due to the region of largest Z500 bias in the North Atlantic that project on the anomalies associated with Atlantic Trough and Atlantic Ridge. The third lowest forecast skill is found for European Blocking, which is known to be particularly challenging to predict compared to other regimes, looking at the categorical weather regime definition (see \citet{Bueler2021}). For the remaining four regimes, a common order of skill cannot be discerned. However, it is noteworthy that for the EMOS-G forecasting method, the CRPSS across all four regimes (Zonal regime, Scandinavian Trough, Scandinavian, and Greenland Blocking) is remarkably similar.
Overall, largest differences in terms of forecast skill for the different forecasting methods and regimes is observed for lead times of 10--14 days (insets in Figures\,\ref{fig:CRPSS}b-\ref{fig:CRPSS}d).

\begin{figure}[!ht]   
\centering
\includegraphics[width=0.4\linewidth]{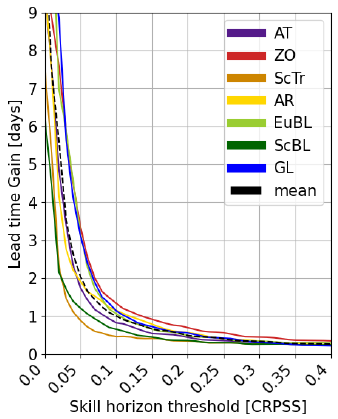}
\caption{Forecast skill horizon gain as a function of different CRPSS thresholds from non-calibrated forecasts to EMOS-G forecasts. Lead time gain separated for weather regimes.}
\label{fig:skillgainCRPSS}
\end{figure}

To asses the skill score improvements of the EMOS-G forecasts in comparison to the non-calibrated forecast in more detail, we calculate the lead time gain across a range of forecast skill horizon thresholds (Figure\,\ref{fig:skillgainCRPSS}). The forecast skill horizon is defined as the lead time at which the CRPSS of a forecast falls below a certain threshold. Typical thresholds for the forecast skill horizon are 0, which indicates that the CRPS of the forecasting method achieves the same score as the climatological reference forecast or 0.1 (see \citet{Bueler2021} in the context of weather regimes), which indicates an improvement of the skill score compared to climatology of 10\%. There is no set rule as to which threshold should be analysed as this is subject to the forecast question. Therefore, we provide a visualisation of a range of thresholds ranging from 0.0 up to 0.4 at intervals of 0.01, demonstrating the robustness of the results across a range of thresholds (Figure\,\ref{fig:skillgainCRPSS}).
EMOS-G consistently outperforms the non-calibrated forecast, as all lead time gain values are positive for each weather regime. The mean lead time gain ranges from 0.3 days for a CRPSS threshold of 0.4 up to 11.2 days for a threshold of 0.0. The most significant improvements occur in forecasts of the Zonal Regime, Greenland Blocking, and European Blocking, while the smallest improvements are observed for forecasts of the Scandinavian Trough and Scandinavian Blocking. This is in line with the close proximity of the CRPSS curves for these four regimes (excluding European Blocking) in Figure,\ref{fig:CRPSS}d. The apparent similarity in terms of forecast skill is due to the substantial increase of forecast skill for the Zonal Regime and Greenland Blocking after being post-processed with EMOS-G. Fixing the CRPSS threshold to 0.1, similar to \citet{Bueler2021}, we find a mean forecast skill horizon of 13.5 and 14.5 days for the non-calibrated and EMOS-G forecast, respectively, indicating a lead time gain of 1 day.

\begin{figure}[!ht]   
\includegraphics[width=1\textwidth]{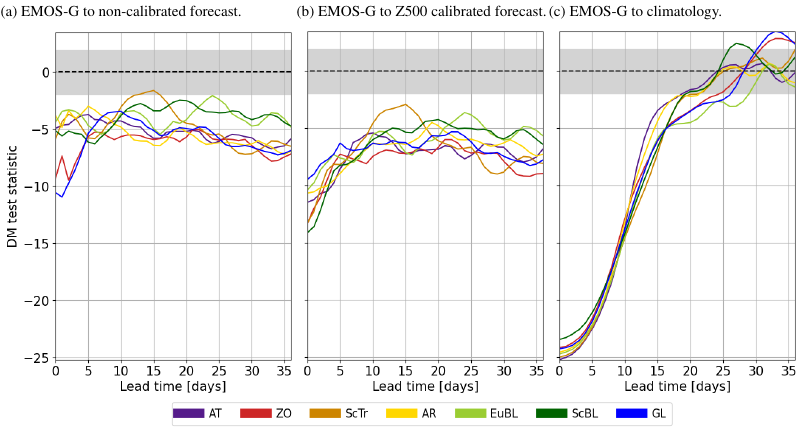}
\caption{Testing the significance of the univariate CRPS improvements of the EMOS-G method with respect to (a) the non-calibrated forecast, (b) the Z500 calibrated forecast and (c) the climatological forecast using a Diebold-Mariano test of equal performance with an $\alpha$-level of 0.05 (indicated by the grey shading).}
\label{fig:Diebold_univariate}
\end{figure}

To complete the univariate analysis, we evaluate the significance of skill improvements using a Diebold-Mariano test for equal performance on the CRPS. We compare the EMOS-G method with the non-calibrated (Figure\,\ref{fig:Diebold_univariate}a) and Z500 calibrated (Figure\,\ref{fig:Diebold_univariate}b) forecasts, as well as with the climatological reference forecast (Figure\,\ref{fig:Diebold_univariate}c).
EMOS-G leads to significantly higher skill than the non-calibrated and Z500 calibrated forecasts. These results are statistically significant for all weather regimes and all lead times at a level of 0.05, except for the Scandinavian Trough at lead times between 12 and 16 days (Figure\,\ref{fig:Diebold_univariate}a). 
The least significant results on the extended lead time when comparing EMOS-G forecasts to non-calibrated forecasts are observed for the European and Scandinavian Blocking, which is in line with the respective verification rank histograms in the supplementary Figures\,\ref{fig:2D_rankhistblocked}d and\,\ref{fig:2D_rankhistblocked}f and Figures\,\ref{fig:2D_rankhistblocked}g and\,\ref{fig:2D_rankhistblocked}i, respectively.
Compared to climatology (Figure\,\ref{fig:Diebold_univariate}c), EMOS-G demonstrates significant performance improvements across all weather regimes out to 19 days forecast lead time, with even longer significant improvements for the European Blocking, reaching out to 28 days.

In conclusion, applying EMOS-G post-processing to the raw, non-calibrated forecasts leads to significant skill improvements across all weather regimes and lead times compared to the method of Z500 calibration. The forecast skill horizon, measured by a 10\% CRPSS improvement relative to climatology, extends to an average of 15.5 days for all weather regimes, surpassing the current practice of Z500 calibration by 1.2 days and the non-calibrated forecasts by 1 day.

\subsection{Multivariate post-processing}\label{sec:multivariate}
Similar to the evaluation of univariate post-processing skill (Section\,\ref{sec:univariate}), we compare skill scores with the non-calibrated and Z500 calibrated forecasts and assess the significance of the skill differences using a Diebold-Mariano test. To evaluate the multivariate skill of the forecasts, we employ the energy score (ES), which is a multivariate extension of the continuous ranked probability score (CRPS). We also evaluate our results by using the variogram score (VS) as an alternative metric which has been argued to be more discriminative with respect to the correlation structure.
%
\begin{figure}[!ht]   
\includegraphics[width=1\textwidth]{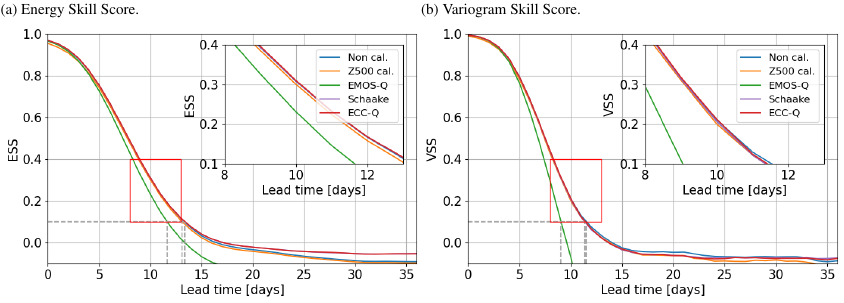}
\caption{Multivariate skill scores for non-calibrated forecast (blue), Z500 calibrated forecast (orange), EMOS-Q (green), Schaake Shuffle (purple) and ECC-Q (red). Using (a) the energy skill score and (b) the variogram skill score.}
\label{fig:multivariateSS}
\end{figure}

When comparing the multivariate skill scores (ESS in Figure\,\ref{fig:multivariateSS}a and VSS in Figure\,\ref{fig:multivariateSS}b) of the univariate post-processing of EMOS-Q (green lines) with the additional multivariate post-processing of EMOS-Q plus ECC-Q (red lines), the necessity of the multivariate step becomes clear. EMOS-Q ensemble members are sorted in ascending order (ensemble member 1 has the lowest values of each weather regime index and ensemble member 11 the highest values), while for ECC-Q the IWR values are sorted based on the rank order of the non-calibrated forecast ensembles. This comparison demonstrates the direct effect of the multivariate post-processing step. The forecast skill improvement of ECC-Q in comparison to EMOS-Q is notable as early as of 5 days lead time. The more relevant comparison of multivariate skill scores is between ECC-Q and non-calibrated and Z500 calibrated forecasts.
The non-calibrated forecasts (blue), Z500 calibrated forecasts (orange) and ECC-Q post-processed forecasts (red) exhibit comparable skill for lead times up to 12 days for ESS and VSS (Figure\,\ref{fig:multivariateSS}a and\,\ref{fig:multivariateSS}b). On extended lead times, the energy skill score for the non-calibrated and Z500 calibrated forecasts is comparable and ECC-Q exhibits higher skill. However, the score of all three forecasting methods is below 0 after lead times of 16--18 days and hence less skilful than a climatological forecast. For the variogram skill score, the results are similar for the non-calibrated and Z500 calibrated forecasts but both exhibit less skill than climatology after 14 days of forecast lead time. The superiority of ECC-Q on the extended lead times does not prevail for the variogram skill score and its performance is similar to the other forecasting methods. In preliminary tests with the Schaake Shuffle, we observed no significant differences in multivariate performance with respect to ECC, based on the ESS and VSS (Figure\,\ref{fig:multivariateSS}, purple line covered by red line). 
%
\begin{figure}[!ht]   
     \centering
     \includegraphics[width=0.5\linewidth]{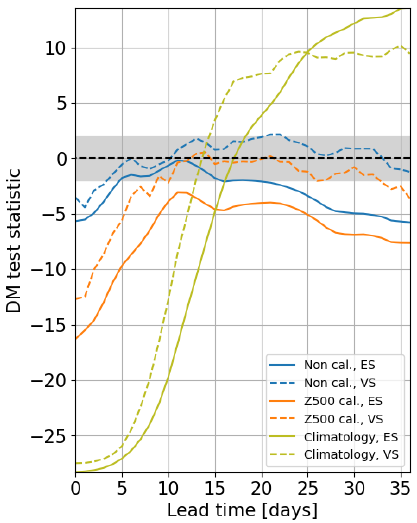}
     \caption{Diebold-Mariano tests based on the multivariate scores (energy score (solid) and variogram score (dashed)) for equal predictive performance of ECC-Q and the non-calibrated forecast (blue), the Z500 calibrated forecast (orange), and the climatological reference forecast (olive green).}
     \label{fig:Diebold_multivariate}
\end{figure}

Analysing the significance of the ECC-Q skill scores against the other forecasting methods, by using a Diebold-Mariano test (Figure\,\ref{fig:Diebold_multivariate}), gives a clearer insight of the actual differences of the skill scores. Using the energy score, ECC-Q does perform better than the non-calibrated and Z500 calibrated forecasts on all lead times. These results are significant for all lead times against the Z500 calibration and significant for all lead times, except day 5--15, against the non-calibrated forecasts. Comparing the energy score of ECC-Q against climatology it becomes apparent that ECC-Q has significant better scores until lead time 16 days. When using the variogram score, ECC-Q performs significantly better up to 8 days of lead time. Comparing against the non-calibrated forecasts, the variogram score of ECC-Q is significantly better until 3 days. Against climatology, the score of ECC-Q is significantly better until lead time 12 days.

In conclusion, the multivariate comparison of the results aligns with the findings from the univariate comparison. Post-processing using EMOS-G and ECC-Q demonstrates its competitiveness with the pre-processing method of Z500 calibration and continuously outperforms it using the energy skill score and for most lead times using the variogram skill score. Restoring the multivariate dependency structure has been shown to be a crucial aspect of the post-processing method, as it leads to a substantial improvement in multivariate performance compared to the univariate post-processing (EMOS-Q), ensuring a more accurate representation of the true relationship between weather regime indices of each ensemble member. However, it is important to acknowledge that the multivariate forecast skill in the extended-range for all forecasting methods, evaluated over the entire testing period, is inferior to climatology.

\subsection{Sensitivity of post-processed weather regime forecasts on training data availability}\label{sec:operational}
The main setup of our analysis involves combining two ECMWF model cycles (Cy46R1 and Cy47R1) to train and evaluate on a large dataset. While this approach allows us to test the potential performance limit of our post-processing method, it does not fully represent the data amount available in an operational forecasting scenario, where training solely relies on reforecast data from the operational model cycle. To address the issue of available reforecast data for training, we test the sensitivity of EMOS+ECC trained on two variants of a reduced set of reforecasts.
\begin{figure}
    \centering
    \includegraphics[width=0.6\linewidth]{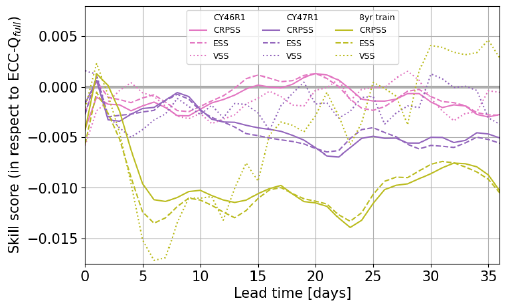}
    \caption{Comparison of performance when reducing training data to the cycles CY46R1 (pink) and CY47R1 (purple), separately, as well as using only the 8 most recent years of both cycles (green) together. We compare the performance of these method setups using skill scores (CRPSS: solid lines, ESS: dashed lines, VSS: dotted lines) with respect to the ECC-Q method trained on both forecast cycles and the full training period of 16 years. Verification is performed on the joined forecasts of both cycles, as in the previous sections.
    }
    \label{fig:small_training}
\end{figure} 

First, we test for a lower initialisation frequency of forecasts by splitting the training data into the respective forecast cycles, Cy46R1 and Cy47R1. Second, we test for the importance of interannual variability in the training period by keeping both forecasting cycles but reducing the amount of training years to the 8 most recent years of the entire training period (June 2007 to May 2015). The testing period is identical to the previous sections, using the 899 forecasts from June 2015 to May 2020, combining data from both forecast model cycles. To ensure a fair comparison, we exclude data from April 26 to July 15 of each year for Cy47R1, as this cycle did not run a full year operationally, hence training EMOS with a 31 day rolling window might not be possible at all or only on a minimal set of forecasts. We now directly compare the performance of the ECC-Q setups with a reduced training period against the ECC-Q setup with the full training set (Figure\,\ref{fig:small_training}) for the skill scores: CRPSS (solid lines), ESS (dashed lines), and VSS (dotted lines).

ECC-Q trained on Cy46R1 (pink), which ran for more than a full year, exhibits skill scores nearly identical when compared to the combination of both forecast cycles (skill score values around 0). ECC-Q trained only on Cy47R1 (purple) exhibits slightly lower skill scores than the ECC-Q trained on Cy46R1, which potentially is due to the unavailability of training data for April 26 to July 15. When training on an 8-year long training period (green) the performance is in general worse than ECC-Q trained only on one forecast cycle and consequently training on the full training data. This indicates that accounting for the interannual variability in the training period is more important than accounting for the initialisation frequency of the forecasts.

In a simplified operational scenario for end-users, it may be beneficial to define a categorical weather regime index by assigning the weather regime with the highest weather regime index (if it is above 1.0) to the corresponding initial date plus lead time (see e.g. \citet{Bueler2021} for a more detailed definition of the categorical weather regime index). With this approach, our results hold when analysing the forecasts using a categorical forecast skill score, namely the Brier skill score (not shown here).

In conclusion, the post-processing approach we evaluated in our study could be directly applied to operational ECMWF extended-range forecasts without considerable losses in skill compared to the results shown above. Training the ECC-Q across many years of forecasts is of greater importance than training the ECC on forecasts with higher initialisation frequency. Our findings remain robust when using a simplified categorical weather regime index instead of the 7-dimensional weather regime index.

%% file: 4discussionconclusion.tex
\section{Conclusions and discussions}\label{sec:Discussion}
The present study explores the potential of a statistical post-processing technique that combines ensemble model output statistics and ensemble copula coupling to enhance the forecast skill of multivariate probabilistic weather regime forecasts. Following the approach of \citet{Grams2017}, we employ a year-round 7-dimensional weather regime index (IWR) that identifies four anticyclonic and three cyclonic regimes. The IWR represents the projection of 500-hPa geopotential height anomalies (Z500A) onto the mean anomaly patterns of the seven distinct weather regimes.

Our approach involves the computation and post-processing of weather regime indices, based on the Z500 field obtained from ECMWF's sub-seasonal reforecast ensemble data, utilising model cycles Cy46R1 and Cy47R1. The outcomes of this process are validated against ERA5 reanalyses. To enhance the accuracy of the raw multivariate probabilistic weather regime forecasts, a combined approach of EMOS and ECC is employed as part of the post-processing procedure.

Biases in the non-calibrated IWR forecasts can be directly traced back to biases in the Z500A fields. EMOS can effectively correct a portion of these biases and systematic forecasting errors, and thus improves univariate forecasting skill scores on all lead times. When evaluating EMOS in an univariate context against the current practice of using Z500 calibrated fields (where Z500A are computed against a model climatology rather than the ERA5 climatology), significant improvements are observed across all lead times and for all weather regimes. The forecast skill horizon, which is defined as the lead time until which the $\textrm{CRPSS}_\textrm{clim}$ exceeds a specific threshold, indicates that EMOS outperforms the current practice and the raw forecast for a range of skill horizon thresholds ($0\leq \textrm{CRPSS}_\textrm{clim}\leq 0.4$). When the threshold is set at 0.1, the mean forecast skill horizon across the seven weather regimes in the EMOS process is 14.5 days. This represents an improvement of 1 day compared to the non-calibrated forecast and 1.2 days compared to the Z500 calibrated forecasts.

The multivariate dependency structure of the IWRs is lost when post-processing ensemble forecasts in an univariate manner with EMOS. Hence, restoring this structure through ECC is crucial in multivariate post-processing. The effectiveness of ECC is evident when comparing multivariate skill scores of the univariate EMOS and multivariate ECC forecasts. The enhancements of ECC compared to EMOS become evident starting from a lead time of 5 days when considering multivariate skill scores (ESS and VSS). Consistent with the univariate comparison against the current practice, the multivariate comparison also demonstrates the superiority of the ECC process. Specifically, ECC significantly outperforms the current practice in terms of the energy score across all lead times and for the variogram score up to a lead time of 8 days.

The EMOS-ECC process exhibits little sensitivity to the initialisation frequency of reforecasts in the training period but displays a more pronounced sensitivity to the interannual variability in the training data. Nonetheless, the skill achieved by EMOS+ECC when trained on a modified training data set surpasses the current practice of calibrating the Z500 forecast field prior to assigning the weather regime index.

In summary, the statistical post-processing approach of combining EMOS+ECC consistently outperforms the current practice of pre-processing the Z500 forecast field. Not only is this approach computationally efficient, but it also provides a compelling alternative due to its ease of implementation and ability to deliver comparable or superior forecasting skill. Additionally, it is versatile, being applicable to both on-the-fly and fixed reforecast configurations. Our findings remain robust when using a limited training data set. Capturing the interannual variability in the training data set is of greater importance than including more frequent initial times. Further, our findings also remain robust using a simplified categorical weather regime index instead of the 7-dimensional weather regime index.

In line with previous studies, such as \citet{Schefzik2013} on pressure, \citet{Scheuerer2015} on wind speed, and \citet{Schefzik2017} on temperature, the EMOS-ECC-Q approach is also on the 7-dimensional weather regime index comparable or superior to other methods. Similar to the findings by \citet{Schefzik2017}, we observe that the individual EMOS-Q ensemble lacks representation of dependence structures (Figure\,\ref{fig:multivariateSS} EMOS-Q vs. ECC-Q), leading to weaknesses in multivariate scores like the energy and variogram score. However, we address this limitation by combining the univariate EMOS-Q with the multivariate ECC-Q post-processing step, effectively restoring the dependency structure and enhancing predictive skill.

Although the EMOS-ECC approach outperforms the current practice of Z500 calibration, it is important to note that on average skilful forecasts of the daily weather regime index are primarily limited to the medium-range, typically up to 15 days. This limitation is dependent on factors like the season, specific weather regime, and the state of the atmosphere. Similar findings were reported by \citet{Bueler2021} using the categorical weather regime index definition based on Z500 calibrated weather regimes and the Brier skill score. The limited skill horizons across all methods largely stem from the intrinsic predictability limit of the atmosphere and model deficiencies. All methods solely rely on Z500 ensemble forecast. The marginal improvements in forecast skill of post-processed probabilistic weather regime indices prompt the question of whether these improvements propagate into downstream applications (e.g., energy or hydrological forecasts) by utilising post-processed forecasts rather than raw forecasts. Further research is needed to address this question.

During specific atmospheric situations (``windows of forecast opportunity''), such as a strong stratospheric polar vortex or specific phases of the Madden-Julian Oscillation \citep{Madden1971}, the forecast skill horizon for weather regimes may extend, as demonstrated by \citet{Bueler2021}. Building on results from studies exploring predictive skill in mid-latitudes and teleconnection patterns \citep{Ferranti2018, Lee2019, Mayer2020}, we believe that enhancing weather regime forecasts in the extended-range can be achieved by incorporating additional information/predictors representing the state of relevant atmospheric modes into our post-processing method. Neural networks are likely to be the most suitable method for the effective implementation of these features \citep{RaspLerch2018,VannitsemEtAl2021}. By introducing neural networks into the post-processing framework, we anticipate not only an improvement in extended-range predictive skill but also the ability to investigate the characteristics of windows of forecasting opportunity through the application of explainable artificial intelligence (explainable AI) methods.
In addition, it is likely that advancing forecast skill into sub-seasonal lead times can be achieved by focusing on skill during windows of forecast opportunity, rather than average skill over all available forecasts. Hence, it is crucial to identify and explore a priori knowledge linked to improved flow-dependent predictability. We hypothesise that evaluating forecasts based on windows of forecast opportunity will be of imminent importance for downstream applications, providing additional information to the user on whether to trust the forecast or not. 

We are currently engaged in optimising neural networks for the effective post-processing of extended-range weather regime forecasts, aiming not only to enhance predictive skill but also to explore - a priori - forecasting windows of opportunity using explainable AI.

%% file: supplements.tex
\renewcommand{\thesection}{S}
\section{Supplementary material}
\setcounter{figure}{0}
\renewcommand\thefigure{\thesection\arabic{figure}}

\begin{figure}[!ht]   
     \includegraphics[width=1.0\linewidth]{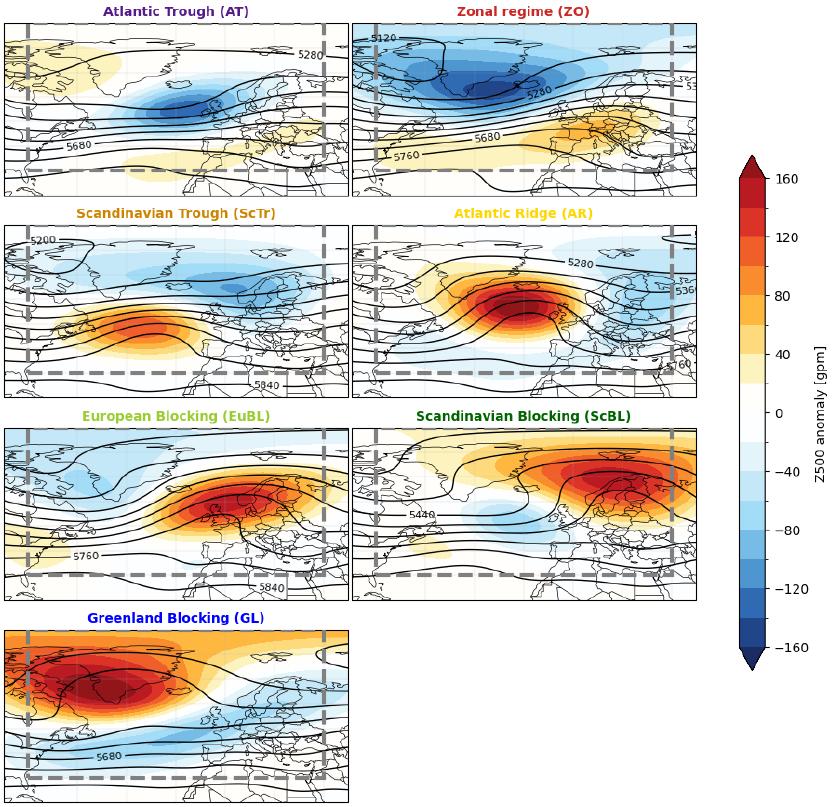}
     \caption{500 hPa geopotential height field (contour) with their anomalies to climatology (shading) for the seven weather regimes: Atlantic Trought (AT), Zonal Regime (ZO), Scandinavian Trough (ScTr), Atlantic Ridge (AR), European Blocking (EuBL), Scandinavian Blocking (ScBL) and Greenland Blocking (GL).}
     \label{fig:WRpatterns}
\end{figure}

\begin{figure}[!ht]   
\includegraphics[width=1\textwidth]{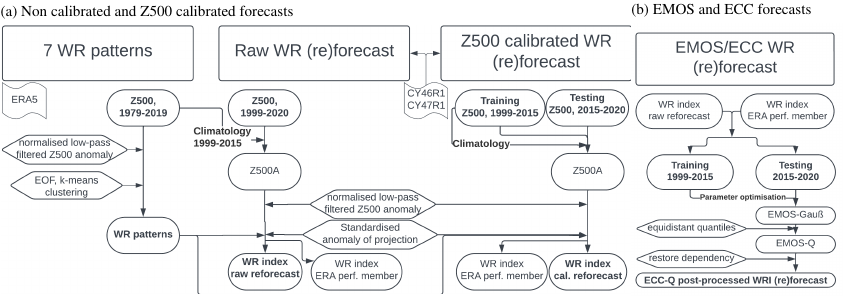}
\caption{Work flow of computing the weather regime index (IWR) of (a) non calibrated and Z500 calibrated forecast and (b) post-processed EMOS/ECC forecasts.}
\label{fig:arrowdiagram}
\end{figure}

\begin{figure}[!ht]   
     \includegraphics[width=1.0\linewidth]{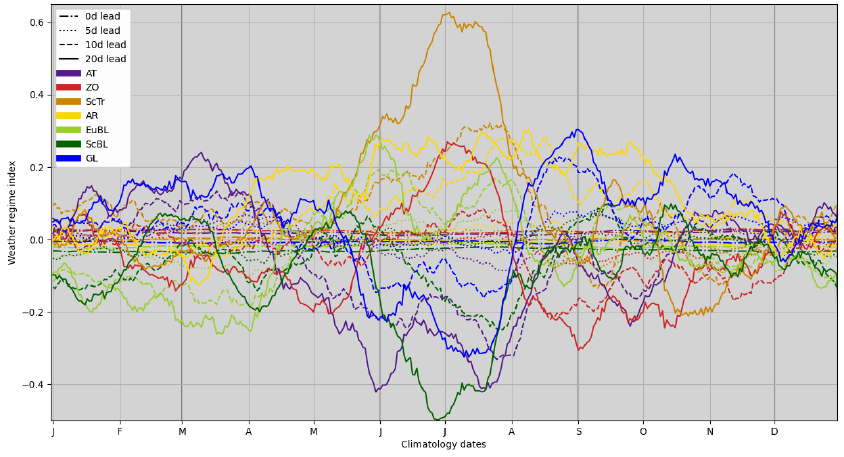}
     \caption{Climatological projection of Z500 biases (see Figure \ref{fig:Z500bias}) onto the weather regime index for 0, 5, 10, and 20 day lead times.}
     \label{fig:IWRbias}
\end{figure}

\begin{figure}[!ht]   
\centering
\includegraphics[width=0.7\textwidth]{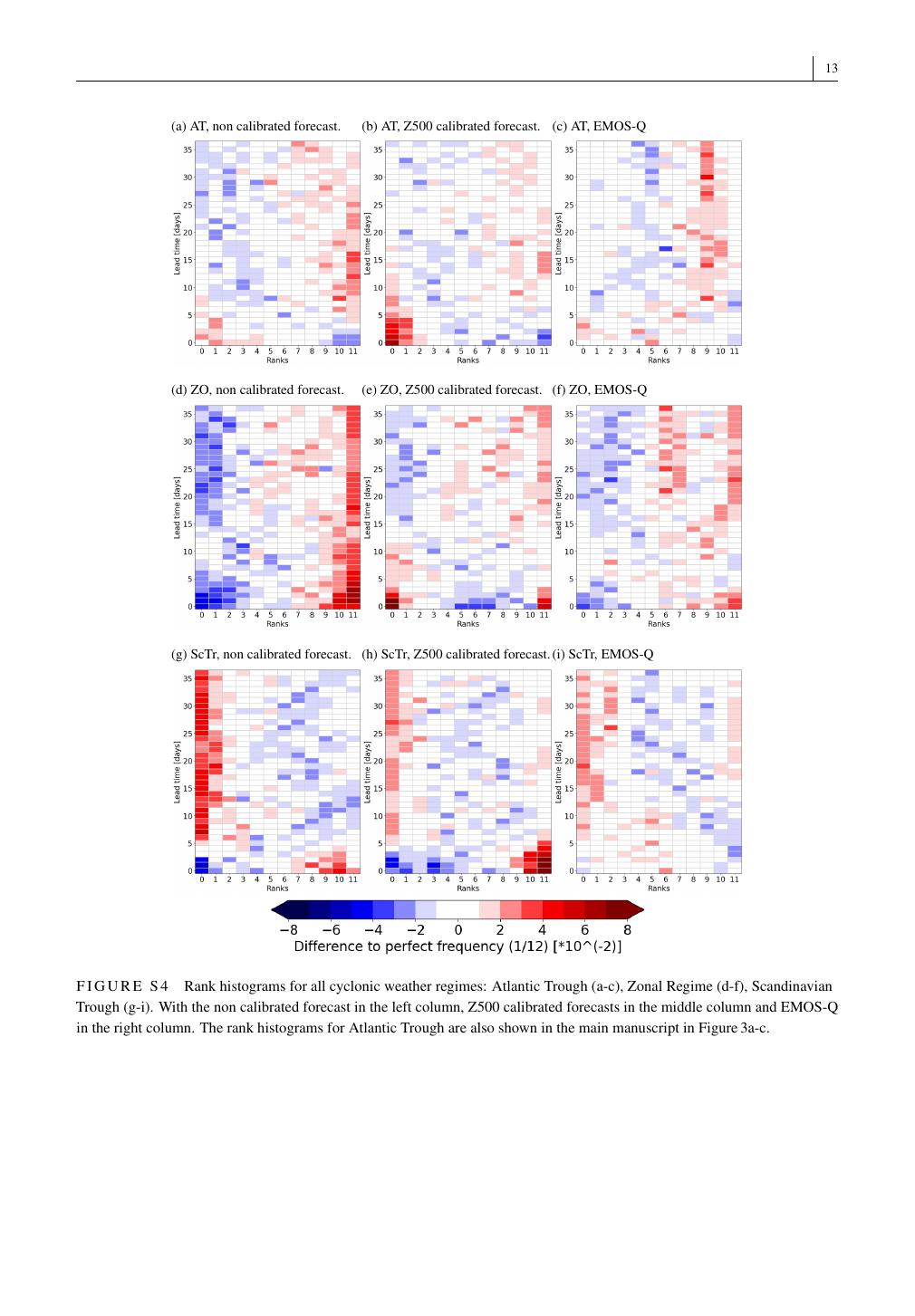}
\caption{Rank histograms for all cyclonic weather regimes: Atlantic Trough (a-c), Zonal Regime (d-f), Scandinavian Trough (g-i). With the non calibrated forecast in the left column, Z500 calibrated forecasts in the middle column and EMOS-Q in the right column. The rank histograms for Atlantic Trough are also shown in the main manuscript in Figure\,\ref{fig:2D_rankhistexamples}a-c.}
\label{fig:2D_rankhistcyclonic}
\end{figure}

\begin{figure}[!ht]   
\centering
\includegraphics[width=0.7\textwidth]{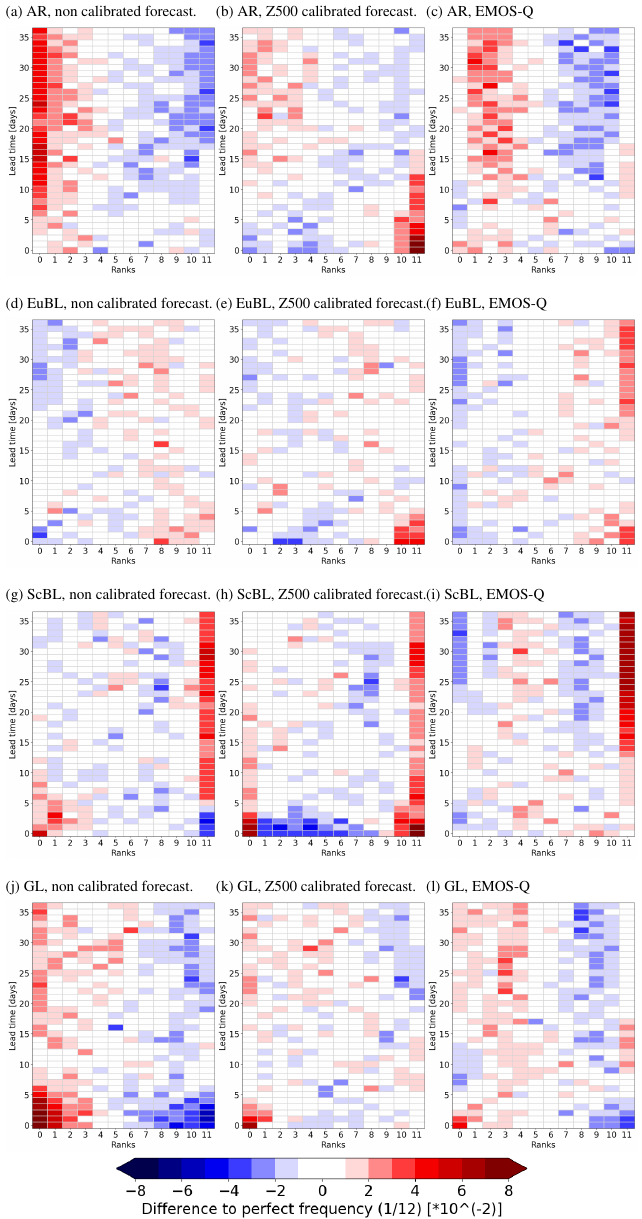}
\caption{Rank histograms for all anticyclonic weather regimes: Atlantic Ridge (a-c), European Blocking (d-f), Scandinavian Blocking (g-i), Greenland Blocking (j-l). With the non calibrated forecast in the left column, Z500 calibrated forecasts in the middle column and EMOS-Q in the right column. The rank histograms for Atlantic Ridge are also shown in the main manuscript in Figure\,\ref{fig:2D_rankhistexamples}d-f.}
\label{fig:2D_rankhistblocked}
\end{figure}